\documentclass{aastex63}

\usepackage{epsfig}
\usepackage{epstopdf}
\usepackage{natbib}
\usepackage{fp}
\usepackage[capbesideposition=right]{floatrow}
\usepackage{wrapfig}

\newcommand\aastex{AAS\TeX}

\newcommand{\Rs}{$\rm R_s$}

\newcommand{\Txi}{$\rm T_{fexi}$}
\newcommand{\Txiv}{$\rm T_{fexiv}$}
\newcommand{\Tc}{$\rm T_{c}$}

\def\ion[#1 #2]{#1\,{\sc #2}}
\def\apj{ApJ}
\def\apjs{ApJS}
\def\jgr{JGR}
\def\solphys{Solar Phys}
\def\insitu{{\it in situ}}

\submitjournal{ApJL}

\shorttitle{\aastex\ Coronal Source Regions of Steady Solar Wind Streams}
\shortauthors{Habbal et al.}

\begin{document}

\title{Identifying the Coronal Source Regions of Solar Wind Streams  from \\
Total Solar Eclipse Observations and {\it in situ}  Measurements Extending Over a Solar Cycle}

\correspondingauthor{Shadia R. Habbal}
\email{shadia@ifa.hawaii.edu}

\author{Shadia R. Habbal}
\affil{Institute for Astronomy,
University of Hawaii,
2680 Woodlawn Drive,
Honolulu, HI, USA}

\author{Miloslav Druckm\"{u}ller}
\affil{Faculty of Mechanical Engineering,
Brno University of Technology,
616 69 Brno, Czech Republic}

\author{Nathalia Alzate}
\affil{NASA, Goddard Space Flight Center, Heliophysics Science Division, Greenbelt, MD, and Universities Space Research Association (USRA), Columbia, MD, USA}

\author{Adalbert Ding}
\affil{Institute of Optics and Atomic Physics, Technische Universit\"at, and Institute for Technical Physics, Berlin, Germany}

\author{Judd Johnson}
\affil{Electricon, Boulder, CO, USA}

\author{Pavel Starha}
\affil{Faculty of Mechanical Engineering,
Brno University of Technology,
616 69 Brno, Czech Republic}

\author{Jana Hoderova}
\affil{Faculty of Mechanical Engineering,
Brno University of Technology,
616 69 Brno, Czech Republic}

\author{Benjamin Boe}
\affil{Institute for Astronomy,
University of Hawaii,
2680 Woodlawn Drive,
Honolulu, HI, USA}

\author{Sage Constantinou}
\affil{Institute for Astronomy,
University of Hawaii,
2680 Woodlawn Drive,
Honolulu, HI, USA}

\author{Martina Arndt}
\affil{Bridgewater State University,
Bridgewater, MA, USA}

\begin{abstract}

This Letter capitalizes on a unique set of total solar eclipse observations, acquired between 2006 and 2020, in white light, \ion[Fe xi] 789.2 nm (\Txi\  = $1.2  \pm 0.1$ MK) and \ion[Fe xiv] 530.3 nm (\Txiv\  = $ 1.8 \pm 0.1$ MK) emission,  complemented by \insitu\ Fe charge state and proton speed measurements from ACE/SWEPAM-SWICS, to identify the source regions of different solar wind streams.  The eclipse observations reveal the ubiquity of open structures, invariably
associated with \ion[Fe xi] emission from $\rm Fe^{10+}$,  hence a constant electron temperature, \Tc\ = \Txi\, in the expanding corona. The \insitu\ Fe charge states are found to cluster around $\rm Fe^{10+}$,   independently of the 300 to 700 km $\rm s^{-1}$ stream speeds, referred to as the continual solar wind. $\rm Fe^{10+}$ thus yields the fiducial link between the continual solar wind and its \Txi\  sources at the Sun. 
While the spatial distribution of  \ion[Fe xiv] emission, from $\rm Fe^{13+}$,  associated with streamers, changes throughout the solar cycle, the  sporadic appearance of charge states $>  \rm Fe^{11+}$, \insitu, exhibits no cycle dependence regardless of speed. These latter streams are conjectured to be released from  hot coronal plasmas at temperatures $\ge \rm $ \Txiv\ within the bulge of streamers and from active regions, driven by the dynamic behavior of prominences magnetically linked to them. 
The discovery of continual streams of slow, intermediate and fast solar wind, characterized by the same \Txi\ in the expanding corona, places new constraints on the physical processes shaping the solar wind.  
 \end{abstract}


\section{Introduction} \label{intro}

To identify the sources of different solar wind streams at the Sun, the topology of expanding coronal structures, as well as the link between their plasma parameters and  corresponding \insitu\ values, need to be established. To that end, the scientific value of simultaneous white light and multi-wavelength observations of emission from forbidden line transitions in the corona, acquired during total solar eclipses, cannot be overstated. These observations remain unique as they capture ubiquitous open structures and their physical properties over a distance range of several solar radii (\Rs), defined here as the expanding corona,  a distance range currently beyond the reach of existing ground- and/or space-based instrumentation. The radial span of this emission is a consequence of the dominance of resonant over collisional excitation for these forbidden lines (see Habbal et al. 2007, 2013). It is within this radial span that a clear separation is established between the plasma that streams freely into interplanetary space, forming the solar wind, and the plasma that remains bound to the Sun such as in active regions and in loop-like structures, seemingly stacked with increasing heights, defining the bulges of streamers. 

Furthermore, coronal emission from the Fe sequence  in the visible and near infrared, namely \ion[Fe ix] to \ion[Fe xiv],  provides a direct inference of the spatial distribution of the electron temperature in the corona (see Habbal et al. 2010a, 2011, Boe et al. 2020a), and hence places constraints on processes responsible for coronal heating and solar wind acceleration. This emission establishes a valuable link between the sources of the emitting Fe ions  in the corona and their distribution in interplanetary space. The choice of Fe as an underlying common parameter to link the corona to the solar wind is obvious: it is the most abundant heavy element in the Sun, which can be easily observed both through the various forbidden emission lines in the corona, as well as \insitu\ via charge state measurements.

S\'ykora (1992a,b) was one of the first to draw attention to the potential connection between \ion[Fe xiv] emission and the observed variability of the solar wind, from ground-based coronagraphic \ion[Fe xiv] observations spanning more than  four solar cycles (see also Ru{\v{s}}in \& Rybansky 2002; Altrock 2004; Badalyan et al. 2005). The first study to quantitatively connect multi-wavelength eclipse observations with {\it in situ} measurements was presented by Habbal et al. (2010b). Using observations from two eclipses, namely 2006 and 2008, and ACE/SWICS Fe charge state measurements (Gloeckler et al. 1998) from 1998 to 2009, these authors discovered the persistent presence of a narrow Fe charge state distribution centered around $\rm Fe^{10+}$, directly linked to \ion[Fe xi] emission in the corona, throughout Solar Cycle (SC) 23.  Lepri \& Zurbuchen (2001) also reported that the average Fe charge state in the ACE data from 1998 to 2000, ranged between Fe$^{9+}$ and Fe$^{11+}$, and that the appearance of much higher charge states was associated with interplanetary coronal mass ejections (ICMEs). More recently, Stakhiv et al. (2016) analyzed ACE charge state measurements from 2007 to 2008 around solar minimum, to search for signatures of the sources of the solar wind. They found that the slow wind has two components, one with properties very similar to the fast steady wind, and another more variable wind. However,  Lepri \& Zurbuchen (2001) and Stakhiv et al. (2016) did not have the coronal observations to link their \insitu\ data back to the Sun.

This Letter capitalizes on the availability of a  complement of co-temporaneous coronal and \insitu\ observations to identify the sources of the solar wind, and the impact of solar activity on them. The coronal observations consist of a unique set of simultaneous white light, \ion[Fe xi] and \ion[Fe xiv] total solar eclipse observations, acquired between 2006 and 2020, from the descending phase of SC 23 to the beginning of SC 25. These data are complemented by  \insitu\ Fe charge state and solar wind speed measurements from ACE, covering the same time period.
We show how this complementary set of observations covering 14 years, yields a link between the prevalence of \ion[Fe xi] emission from $\rm Fe^{10+}$, characterized by \Txi\ = $ 1.2  \pm 0.1$ MK, in the expanding corona, and the \insitu\ presence of {\it continual} solar wind streams, clustering around $\rm Fe^{10+}$, with speeds ranging from $\approx$ 300 to 700 $\rm km s^{-1}$.  These observations also suggest that the sporadic appearance of high Fe charge states \insitu\ can be attributed to the dynamics of prominences at the base of streamers, driving CMEs as well as the more variable solar wind streams.
 
\section{The Data: Total Solar Eclipse Observations  and  \insitu\ Measurements \\
Between 2006 and 2020} \label{obs}

\subsection{High Resolution White Light and Multi-wavelength Total Solar Eclipse Observations}

\begin{figure}[h]
\centering
{\includegraphics[width=0.8\textwidth]{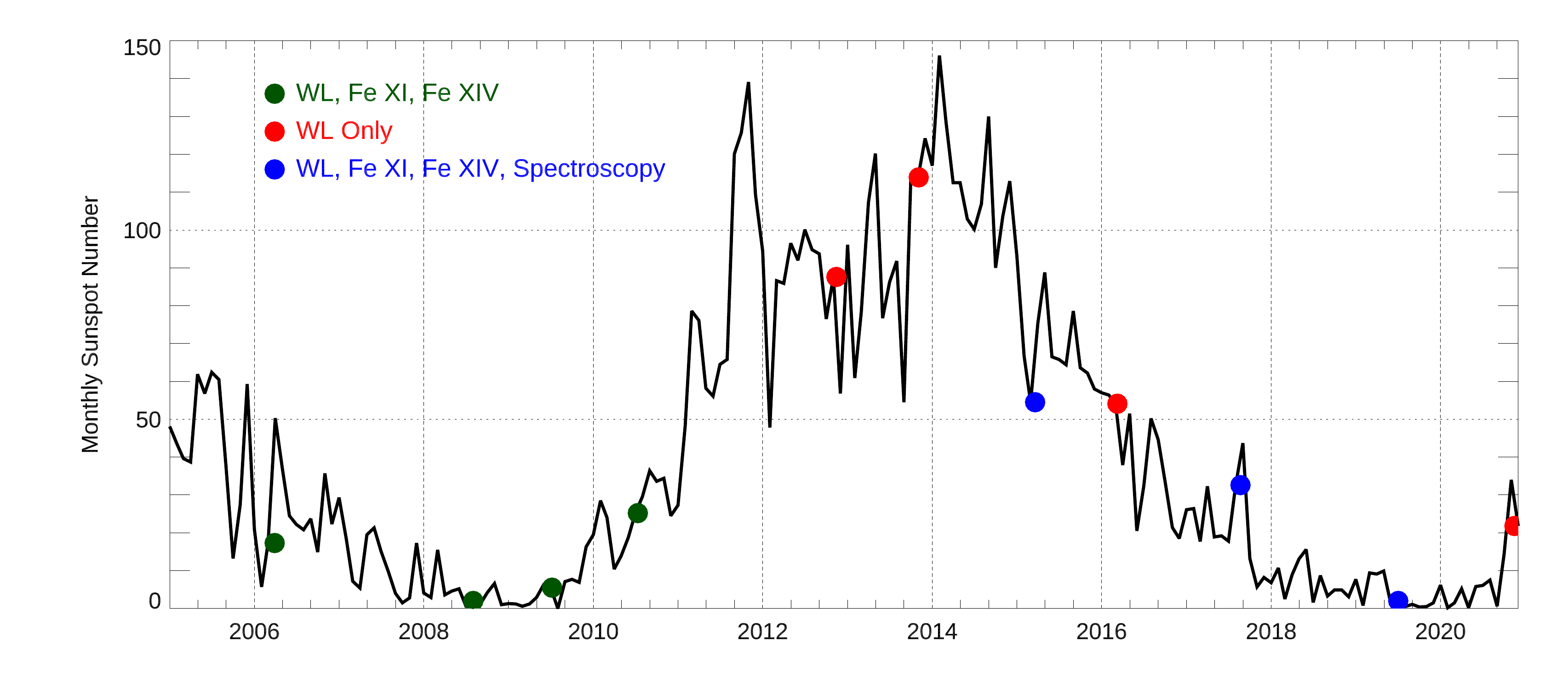}
}
\caption{
Plot of the monthly sunspot numbers with the dates of the eclipse observations given in Table 1 shown as colored circles. Green circles refer to dates with white light (WL), \ion[Fe xi] and \ion[Fe xiv] observations, red is for observations with WL only, and blue is same as green with the addition of spectroscopy. 
}
\label{dates}
\end{figure}

Details of the total solar eclipse observations in white light and Fe coronal emission lines, acquired between 2006 and 2020, are given in chronological order in Table 1, together with the corresponding Carrington Rotation (CR) and SC numbers. These observations are placed in the context of the monthly sunspot number with the dates of the eclipse observations given by colored circles shown Fig. \ref{dates}. These observations straddle 3 cycles, and coincide with different phases of solar activity. The 2006 eclipse occurred during the descending phase of SC 23, with its minimum at 2009. SC 24 had 2 peaks in 2012 and 2013. The 2019 eclipse observations were acquired shortly before the minimum of SC 24 in December 2019.  The 14 December 2020 was acquired at the beginning of SC 25. Since we were clouded out during the 2012, 2013, 2016 and 2020 eclipses, we used white light observations available from amateurs for those years. Unfortunately, since no other teams acquire multi-wavelength observations comparable to ours, such observations are missing from these years.  

High resolution white light images were taken with commercially available large format digital cameras, such as Nikon and Canon, retrofitted with different focal length lenses, using a sequence of exposure times. These images were processed by M. Druckm\"uller who developed mathematical methods for the precise registration of images,  and for the visualization of  coronal structures using adaptive filters inspired by human vision (see Druckm\"uller 2009, 2013; Druckm\"uller et al. 2006).

Details of the instrumentation used for the multi-wavelength Fe emission line observations  and of the data analysis techniques can be found in several publications (e.g. Habbal et al. 2011, Boe et al. 2020a). To summarize, observations for each coronal emission line are acquired with a pair of optical systems retrofitted with 0.5 nm narrow bandpass filters. One unit in the pair is centered at the wavelength of the spectral line, and the other at 1 \--\ 3 nm to the blue. This choice is dictated by the fact that spectral line intensities are only several percents of the intensity of the continuum in the 1 \--\ 3 \Rs\ distance range. To isolate the spectral line emission, the pair is operated simultaneously with the same sequence of exposure times. The data are corrected for dark and flat field exposures acquired after totality. They are then subtracted from each other. As a bonus, observations of the continuum at multiple wavelengths also yield scientifically valuable data pertaining to the properties of the F and K corona, as recently presented by Boe et al. (2021).  

\begin{table}[]
\caption{ Eclipse dates with corresponding observing sites, observers, Carrington Rotation (CR) and Sunspot Cycle (SC).}
\footnotesize
\begin{tabular}{|l|l|l|l|}
\hline
ECLIPSE DATE & TIME (UT) & OBSERVING SITES  \& OBSERVERS & CR/{\bf SC} \\
\hline
\hline
2006 March 29   & 10:13:57 \--\ 10:18:03 & Southern Sahara, Libya (Habbal \& Co.)  & 2041/{\bf 23} \\
\hline
2008 August 1 & 11:03:35 \--\ 11:05:39 & Bor Uzur, Gobi Desert, Western Mongolia (Druckm\"uller \& Co.) &\\  
&  11:13:17 \--\ 11:15:06 & AlShan, Gobi Desert, Western China (Habbal \& Co.)  & 2072 \\
\hline
2009 July 7  &  03:28:39 \--\ 03:34:20 & Enewetak, Marshall Islands (Habbal \& Co.) & 2085  \\
\hline
2010 July 11  & 18:45:36 \--\ 18:50:05 & Tatakoto, French Polynesia (Habbal \& Co.) & 2098/{\bf 24}  \\
\hline
2012 November 13{$^*$}  &  20:37:41 \--\ 20:39:41 & Queensland, Australia (David Finlay \& Constantinos Emmanoulidis) & 2130  \\
\hline
2013 November 3{$^*$}  & 13:52:40 \--\ 13:53:45 & Lamberene, Gabon (Constantinos Emmanoulidis) & 2143  \\
\hline
2015 March 20{$^{++}$}  &  10:10:40 \--\ 10:13:08 & Longyearbyen, Svalbard  (Habbal \& Co.) & 2161\\
\hline
2016 March 9{$^*$}  &  00:22:11 \--\ 00:24:03 & Penyak Beach, Banka Island, Indonesia (Don Sabers, Ron Royer) & 2176  \\
\hline
2017 August 21{$^{++}$}  & 17:21:11 \--\ 17:23:14 & Mitchell, Oregon (Habbal \& Co.) & 2194  \\
\hline
2019 July 2{$^{++}$}  &  20:40:03 \--\ 20:42:16 & Rodeo, Argentina  (Habbal \& Co.) & 2219  \\
\hline
2020 December 14{$^*$}  & 16:07:59 \--\ 16:10:04 &  Fortin Nogueira, Neuquen, Argentina (Andreas M\"oller)  & 2238/{\bf 25}  \\
& 16:21:11 \--\ 16:23:20 &  Bahia Creek,  Rio Negro,  Argentina  (Dario Harari) &  \\
\hline
\end{tabular}
\end{table}
\label{eclipse}

\subsection{The \insitu\ ACE data}

To connect coronal structures to different solar wind streams, we  complement the eclipse observations with the most comprehensive Fe charge state \insitu\ data from 2006 to the end of 2020. These are available from the Solar Wind Electron, Proton, and Alpha Monitor (SWEPAM; McComas et al. 1998) and the Solar Wind Ion Composition Spectrometer (SWICS; Gloeckler et al. 1998) instruments onboard the Advanced Composition Explorer (ACE) spacecraft (Stone et al. 1998),  which orbits around the Sun-Earth L1 Lagrangian point. (See http://www.srl.caltech.edu/ACE/ASC/index.html.)  For the purpose of this study,  we use the solar wind bulk velocities from ACE/SWEPAM, and the Fe charge states from SWEPAM-SWICS, available in 12-minute averages.  The ACE/SWICS data are available in two datasets corresponding to two different instruments, SWICS 1.1 and SWICS 2.0.  The convenience of using this dataset is that gaps in SWEPAM data are filled with SWICS data, when available.  An alteration in the instrument's operational state due to radiation and age resulted in the data gap between 2012 and 2013.

\section{The Case for \ion[Fe xi] and \ion[Fe xiv]}

\begin{figure}[h]
\centering
{\includegraphics[width=1.0\textwidth]{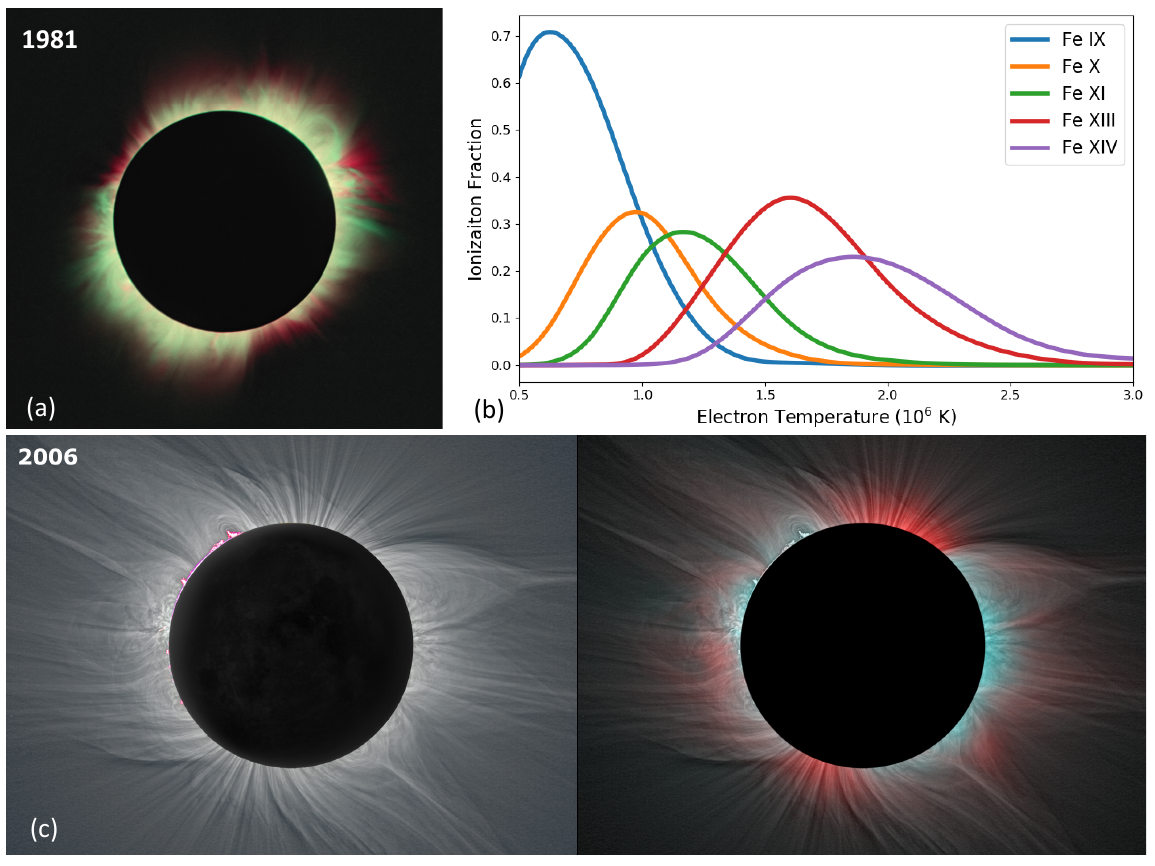}
}
\caption{(a) Composite image of emission in continuum, \ion[Fe x] 637.4 nm (red) with a 0.3 nm bandpass filter, and \ion[Fe xiv] 530.3 nm (green) with a 0.2 nm bandpass filter, from the 1981 eclipse taken with a 130/1950 mm refractor, and recorded on black and white high speed, high dynamic range, Fomapan N800 film. The image was constructed and processed by M. Druckm\"uller.  (b) Ionization fraction curves for the dominant coronal forbidden spectral lines and corresponding charge states, namely:  \ion[Fe ix] (${\rm Fe}^{8+}$),  \ion[Fe x] (${\rm Fe}^{9+}$),  \ion[Fe xi] (${\rm Fe}^{10+}$), \ion[Fe xiii] (${\rm Fe}^{12+}$), and \ion[Fe xiv] (${\rm Fe}^{13+}$) (from Boe et al. 2018). Note that \ion[Fe xii] is absent in this sequence because its two emission lines at 303.3 and 356.1 nm,  are in the near UV, which makes them more challenging to observe from the ground. (c) Eclipse composite image from the 2006 March 29 eclipse, with white light given in the left panel, and a composite of white light, \ion[Fe xi] (red), and \ion[Fe xiii] 1074.7 nm (cyan) on the right.}
\label{ion}
\end{figure}

The first total solar eclipse observation to include {\it simultaneous} continuum and multiwavelength imaging with narrow bandpass filters  for the \ion[Fe x] 637.4 nm and \ion[Fe xiv] 530.3 nm lines, was acquired by the Slovak expedition led by J. S\'ykora to Siberia in 1981, very close to solar maximum. Shown in Fig. \ref{ion}a, this
composite image captures the clear distinction between emission from two spectral lines with different thermal properties. (As an aside, we note that this distinction was first reported by Mitchell, 1932, albeit without his knowledge of the ions associated with these two wavelengths, at that time.)

The distinction between the thermal properties of coronal forbidden lines is best demonstrated by the temperature dependence of the ionization fraction of the suite of \ion[Fe ix] to \ion [Fe xiv] lines, in the visible to the near infrared. Shown in Fig. \ref{ion}b, the curves are calculated using data from Arnaud and Raymond (1992) with no assumption of collisional effects (see Boe et al. 2020a). 
These calculations yield peak ionization temperatures $ \approx$ 0.7, 1.0, 1.2, 1.7 and 1.8 MK, respectively. Their precise values can differ slightly from calculations using the more recent CHIANTI database (Dere et al. 2019), which yield 0.76, 1.05, 1.29, 1.74 and 1.95 MK respectively (Landi, private communication). Consequently,
we assume an uncertainty of 0.1 MK for these values in this work.

 The curves in Fig. \ref{ion}b demonstrate that (1) over the 0.5 \--\  3~$\times~10^6$ K range, different  temperature plasmas can be readily targeted with observations in this sequence of Fe lines, and (2) there exist line pairs that enable the clear distinction between different temperature structures, such as \ion[Fe xi] and \ion[Fe xiv]. 
The first  \ion[Fe xi] 789.2 nm image of the corona, shown in red in the right panel of Fig. \ref{ion}c, (cyan is from \ion[Fe xiii]), was acquired during the 2006 March 29 total solar eclipse (Habbal et al. 2007, 2013). The unexpected spatial extent of the \ion[Fe xi] emission led to the realization that emission from coronal forbidden lines is dominated by radiative excitation once collisional excitation diminishes significantly close to the Sun (Habbal et al. 2007, 2013). With a dependence of the emission on the ion number density, and not on the density squared, characteristic of collisionally-excited extreme ultraviolet lines, emission from coronal forbidden lines can thus be detected out to much larger distances from the Sun.

The first simultaneous observations of the full  Fe suite of \ion [Fe ix], \ion [Fe x], \ion [Fe xi], \ion [Fe xiii], and \ion [Fe xiv], were acquired by Habbal et al. (2011) during the total solar eclipse of 2010 July 11. An image in \ion [Ni xv]  with a peak at 2.5 MK was also taken at that time, to expand the temperature coverage, since \ion[Fe xv]  at 705.86 nm would be contaminated by a telluric absorption line. These observations showed that  \ion[Fe ix] emission was extremely weak, while \ion [Ni xv] emission, which was limited to the bulges of streamers and active regions, was structureless compared to \ion [Fe xiv] emission. They thus demonstrated that the best candidates for investigating the thermal properties of the corona are \ion[Fe x], \ion[Fe xi], \ion[Fe xiii] and \ion[Fe xiv]. 

\begin{figure}[h]
\centering
{\includegraphics[width=0.9\textwidth]{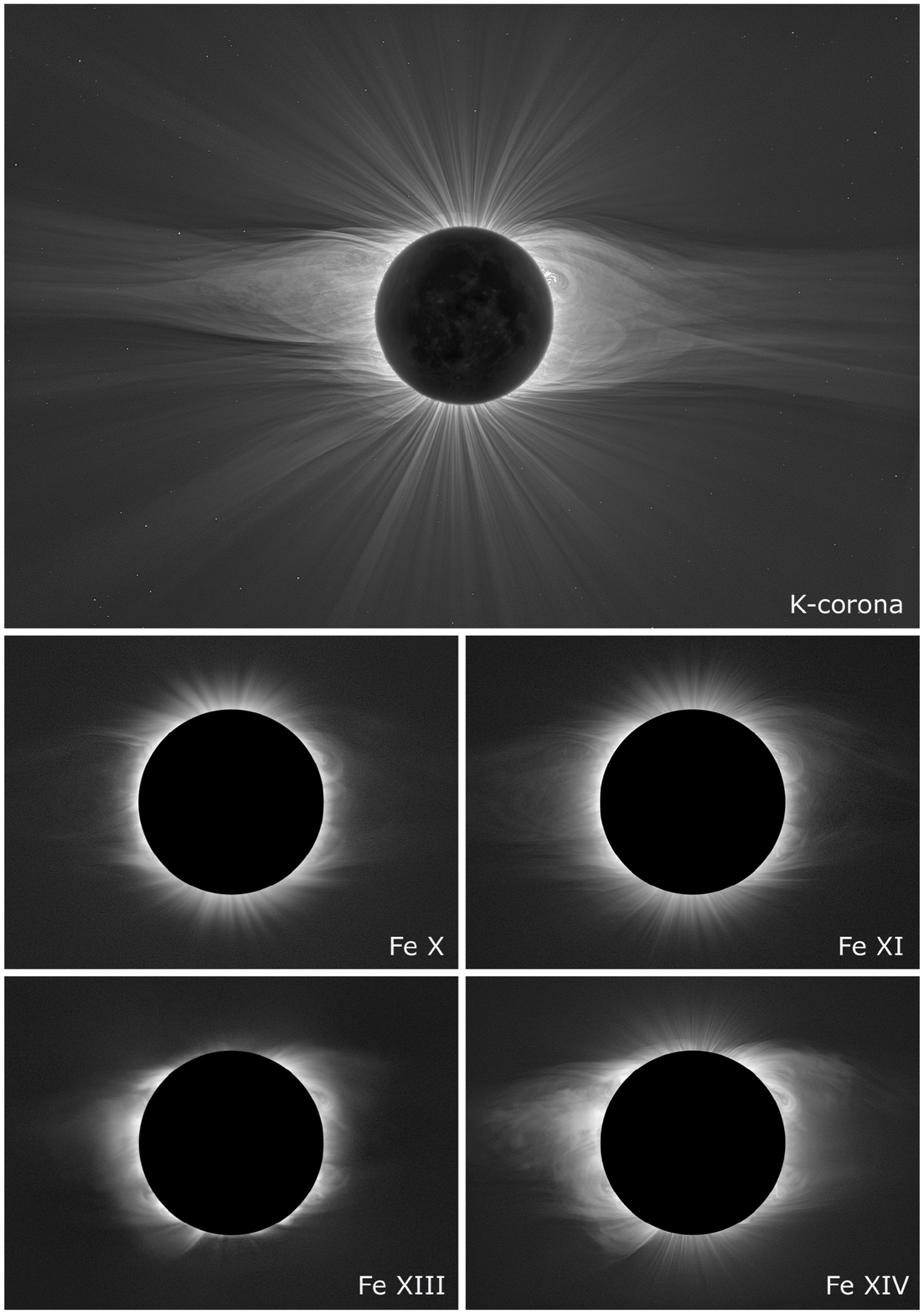}
}
\caption{Comparison of white light,  \ion[Fe x],  \ion[Fe xi], \ion[Fe xiii] and \ion[Fe xiv] emission from the 2019 total solar eclipse observations. Solar north is vertically up.
 }
\label{2019_all}
\end{figure}

It is clear from Fig. \ref{ion}, that imaging in either \ion[Fe x] or \ion[Fe xi] can be used to map the spatial distribution of the `cold' ($\approx$ 1 MK) coronal structures. However,  the higher abundance of $\rm Fe^{10+}$ (i.e. \ion[Fe xi] emission) in the corona (see, Habbal et al. 2010b), compared to $\rm Fe^{9+}$ (i.e. \ion[Fe x] emission), favors the use of \ion[Fe xi]. 
Further supporting evidence for the choice of \ion[Fe xi] over \ion[Fe x],  is provided by the 2019 July 2 total solar eclipse observations shown in Fig. \ref{2019_all}, where the spatial extent of \ion[Fe xi] emission exceeds that of \ion[Fe x]. The comparison between \ion[Fe x] and \ion[Fe xi] in this figure further points to the fact that despite the proximity of their peak ionization temperatures, there are differences in coronal structures between these two lines. This implies that the thermal properties of coronal structures are distinguishable to within $<$ 0.2 MK. 

For the hotter coronal structures, either \ion[Fe xiii] or \ion[Fe xiv] can be used to map the `hot' ($\approx$ 2 MK) structures. However,  \ion[Fe xiv] is the clear winner for two reasons: (1) The continuum solar disk intensity at 530 nm is approximately three times stronger than at 1074 nm. Hence the radiative excitation of these two lines from the solar disk radiation enables the coronal structures in \ion[Fe xiv] to be visible to much larger distances than for \ion[Fe xiii]. (2) Standard CCD and CMOS cameras, such as the ones we use, have a very low quantum efficiency (QE) of $\approx 2 \--\ 5 \% $, at 1000 nm. This choice is further corroborated by the example of Fig. \ref{2019_all}, as it demonstrates why \ion[Fe xiv], whose spatial extent is significantly larger than that of \ion[Fe xiii], is a much better choice. 

It thus becomes clear that  \ion[Fe xi] and \ion[Fe xiv] emission, with their distinct temperature characteristics, remain the strongest among the Fe line sequence, thus dictating the choice of these two emission lines for the work presented here. Furthermore, given the clear distinction with their neighboring lines, i.e. \ion[Fe x] for \ion[Fe xi] and \ion[Fe xiii] for \ion[Fe xiv], we argue that  \ion[Fe xi]  and \ion[Fe xiv] represent two relatively narrow temperature ranges, namely 
 \Txi\ = $1.2  \pm 0.1 $ and  \Txiv\  = $1.8  \pm 0.1  $  MK, respectively.

\section{Changes in the Spatial Distribution of Open Fine-Scale Structures and of the Electron Temperature }
\label{te}

\begin{figure}[h]
\centering
{\includegraphics[width=1.0\textwidth]{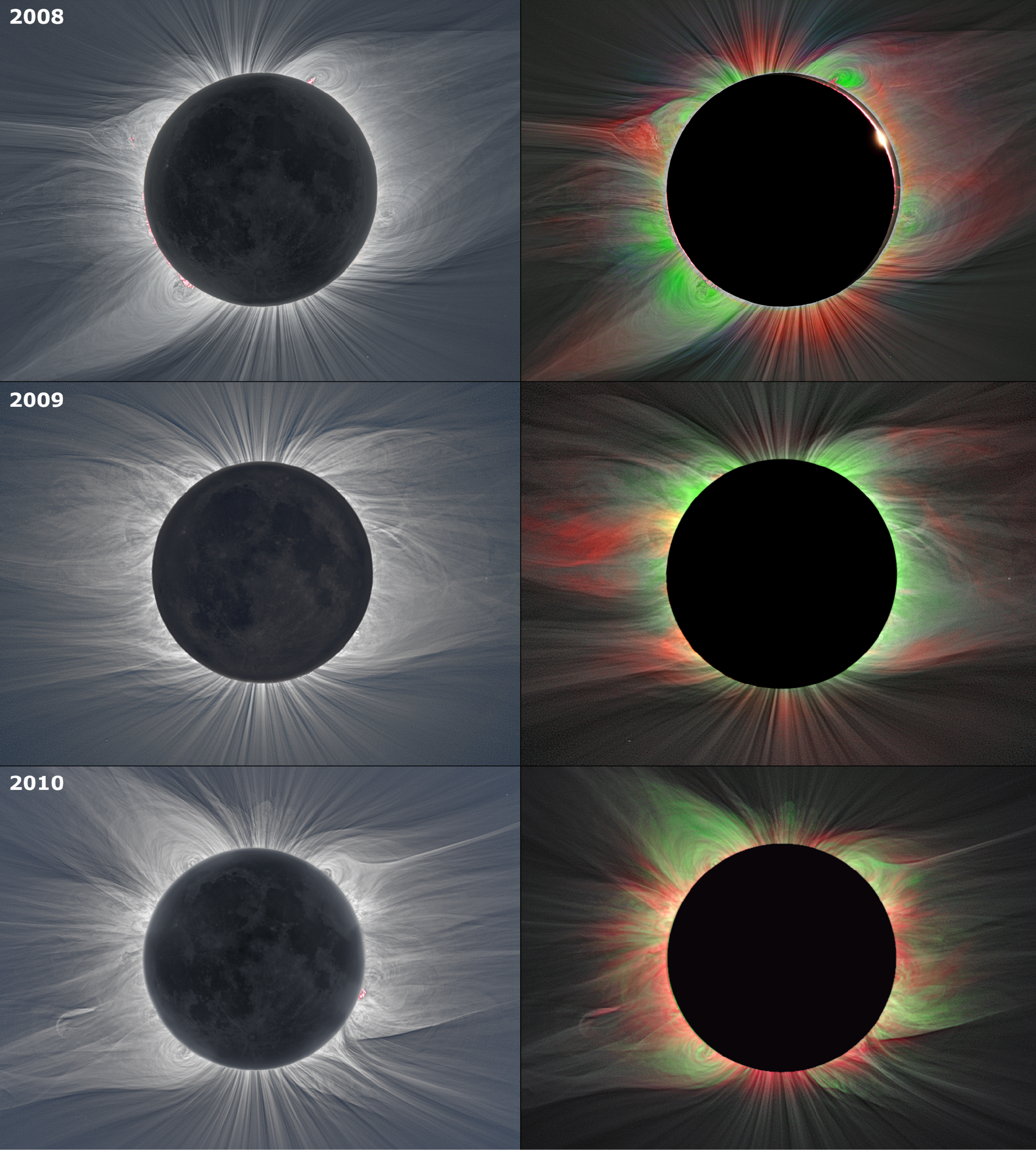}
}
\caption{Composite from total solar eclipse images acquired between 2008 and 2010. To maximize the clarity of the images, they have been split into two panels. As in Fig.\ref{ion}c, the white light image is given in the left column, and the overlay of white light with \ion[Fe xi] 789.2 nm (red) and \ion[Fe xiv] 530.3 nm (green) in the right column. Solar north is approximately vertically up.
 }
\label{allfe1}
\end{figure}

\begin{figure}[h]
\centering
{\includegraphics[width=1.0\textwidth]{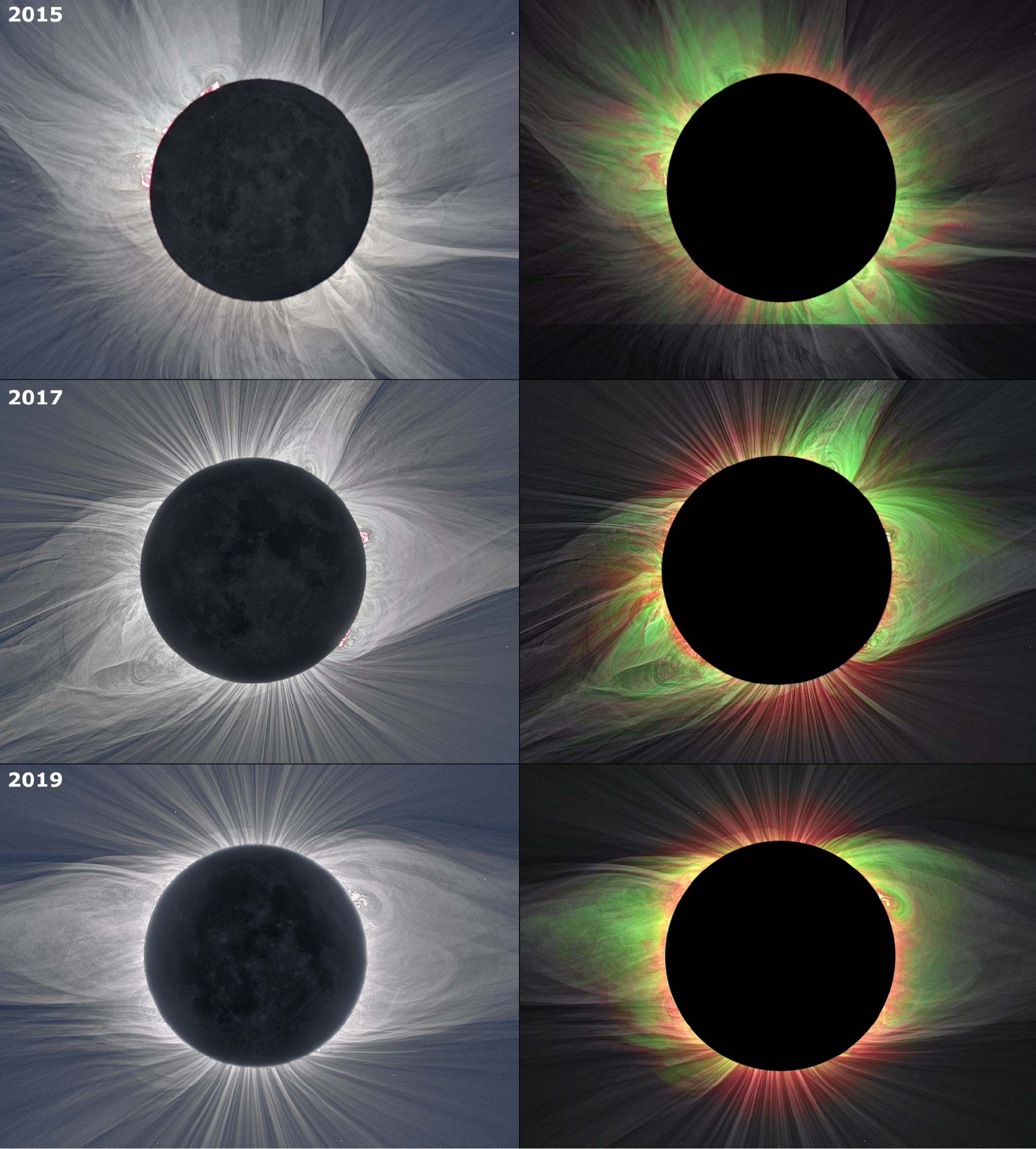}
}
\caption{Same as Fig. \ref{allfe1} for images acquired between 2015 and 2019. The gaps in emission line observations in 2012, 2013 and 2016 were due to poor observing conditions. }
\label{allfe2}
\end{figure}

All \ion[Fe xi] and \ion[Fe xiv] total solar eclipse observations acquired simultaneously with 0.5 nm narrow bandpass filters, are shown in Fig. \ref{ion}c,  \ref{allfe1} and \ref{allfe2}. The layout in these figures is such that white light is given in the left panels, and the \ion[Fe xi] (red) and \ion[Fe xiv] (green) composites, together with white light, are shown in the right panels (except in Fig. \ref{ion}c, where \ion[Fe xiii] in cyan replaces \ion[Fe xiv]). Although both panels show the same field of view, we note that white light can be visible out to at least 15 \--\ 20 $\rm R_s$ during eclipses, while the signal to noise ratio for the \ion[Fe xi] and \ion[Fe xiv] intensities becomes too weak to detect beyond 3 \Rs. While white light visualizes all coronal structures, independently of their temperature, a comprehensive visual representation of the thermal structure of coronal magnetic fields emerges when white light is combined with the distribution of \ion[Fe xi] and \ion[Fe xiv] emission, as is evident in the right panels.

In the composite images, \ion[Fe xiv] emission is found to be invariably linked to streamers, while \ion[Fe xi] emission is found to be unequivocally associated with all open field lines, throughout the corona. The latter also seems to often infiltrate streamers in the plane of the sky, underscoring the filamentary nature of all coronal structures and their low filling factor. 

More importantly, the span of these observations over more than a full solar cycle enables the investigation of the impact of solar activity not only on the topology of coronal magnetic fields, but also on the temperature distribution in the corona. The temporal coverage from 2006 to 2020 shows that the spatial distribution of \ion[Fe xiv] emission, characterized by \Txiv\ changes across the corona as a function of time within a cycle. This change can lead to the unverified impression that the corona becomes hotter with increased solar activity. On the other hand, the ubiquity of the \ion[Fe xi] emission persists regardless of solar activity,  implying that the expanding corona is constrained to approximately \Txi\ regardless of phase within a solar cycle.

In summary, the spatial distribution and radial extent of the emission from \ion[Fe xi] and \ion[Fe xiv] in these composite images spanning more than a solar cycle, demonstrate, for the first time, how the distribution of the hotter \ion[Fe xiv] coronal emission at \Txiv, changes throughout the different phases of a solar cycle, while the cooler \ion[Fe xi] emission at \Txi, remains spatially ubiquitous throughout the expanding corona, independently of phase within a cycle.

\section{Manifestations and Sources of Dynamic Events in the Corona}

Despite the very short duration of totality, eclipse images, whether in white light or coronal emission lines, capture not only the `static' state of the corona, but also its instantaneous dynamic status. Manifestations of dynamic events, such as CMEs, waves and turbulence, are described next. It is shown how the distinct complex prominence structures are intricately connected to surrounding arch-like structures, and are the most likely drivers of the variable solar wind measured \insitu. 

\subsection{CMEs, Waves and Turbulence}
\begin{figure}[h]
\centering
{\includegraphics[width=0.8\textwidth]{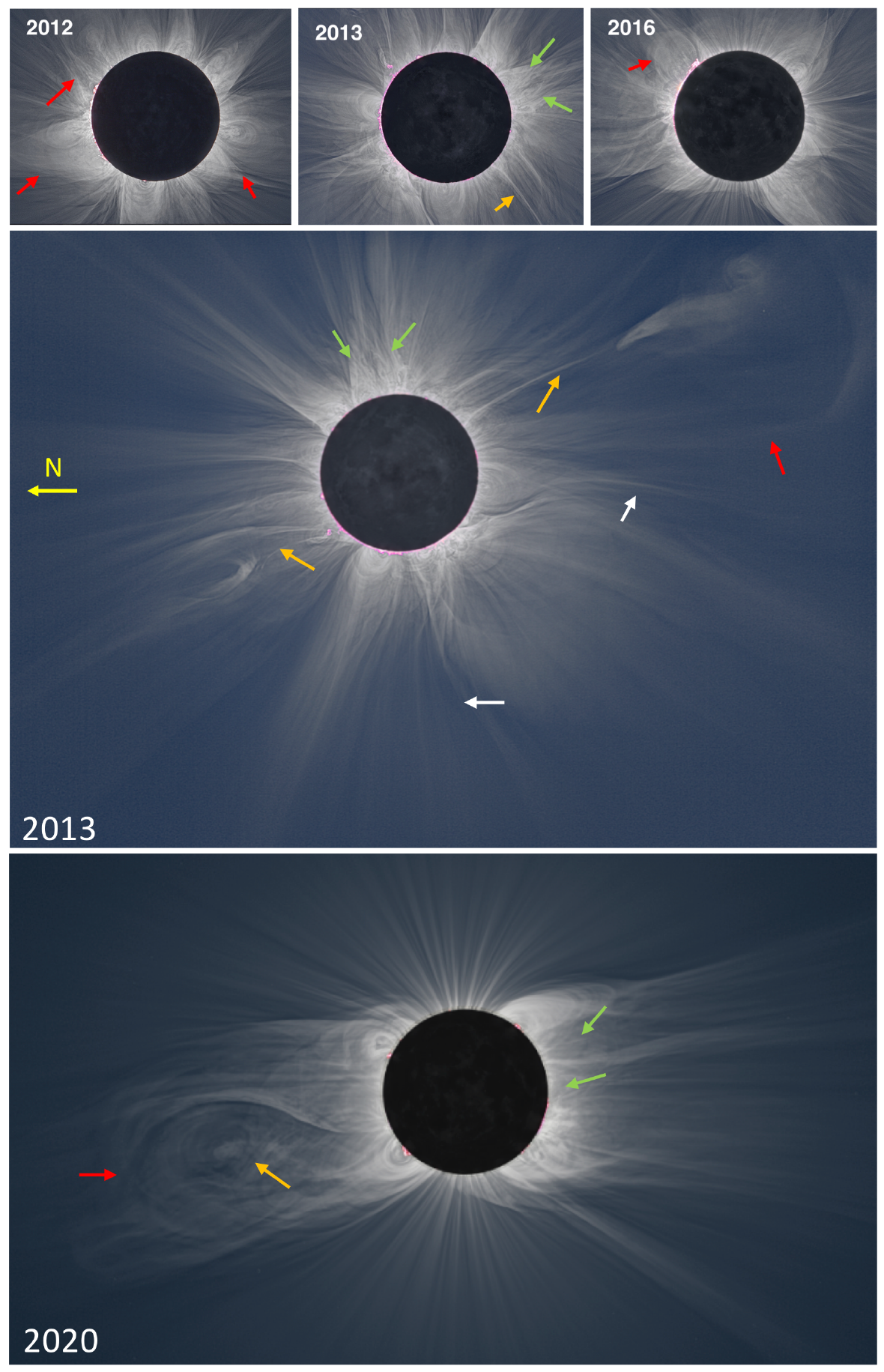}
}
\caption{Top row: White light images from 2012, 2013 and 2016.  Middle:  Larger field of view of the 2013 white light image showing the full extent of a bulb-like CME envelope.  Bottom: White light eclipse image from 2020 with a CME bubble. In all panels, red arrows point to CMEs, either at their emergence very close to the Sun as in 2012 and 2016, or their full extent as in 2013 and 2020.  Green arrows point to wavy motions. Orange arrows point to prominence cores embedded within a CME. For 2013, the white arrows point to the bounding imprints of the passage of a CME. (Solar north is approximately vertically up in all panels except for the larger 2013 field of view where it points to the left).
}
\label{allWL} 
\end{figure}

 Fig. \ref{allWL} is a compilation of the white light eclipse images from 2012, 2013, 2016 (top panels) and 2020, for which there were no corresponding coronal emission line images. This figure provides a comprehensive overview of dynamic structures, such as CMEs (red arrows), with their full extent captured in the wider fields of view of the 2013 and 2020 eclipses shown in the lower panels.  In the 2013 wide field of view image, the faint boundary of the full CME bubble is shown by the red arrow. The prominence ejecta forming its core as well as that of another CME, almost diametrically opposite, are identified by the orange arrows.  The two white arrows point to the imprint of a CME that had passed through the corona prior to the eclipse time (see Alzate et al.  2017,  and Druckm\"uller at al. 2017, for more details). The green arrows point to wavy patterns. 
A full CME was also captured off the east limb in 2020 (red arrow), shown in the lower panel. The twisted features of the prominence ejecta (orange arrow) at its core, are clearly evident. Wavy patterns  (green arrows) are present off the west limb, as in 2013 and in 2020. 

Missing from these images is the temperature characteristic associated with the dynamic events. This information is however available in previous eclipses, such as in 2017 and 2019 (see Fig. \ref{allfe2}). In 2017, the complex structure of the CME in the south east was clearly dominated by \ion[Fe xiv] emission, as found in streamers. In 2019, the bulge of the east streamer was dominated by large-scale eddies, also seen in \ion[Fe xiv] emission. 

\subsection{Prominence-Corona Connectivity} 
\label{proms}

\begin{figure}[h]
\centering
{\includegraphics[width=0.35\textwidth]{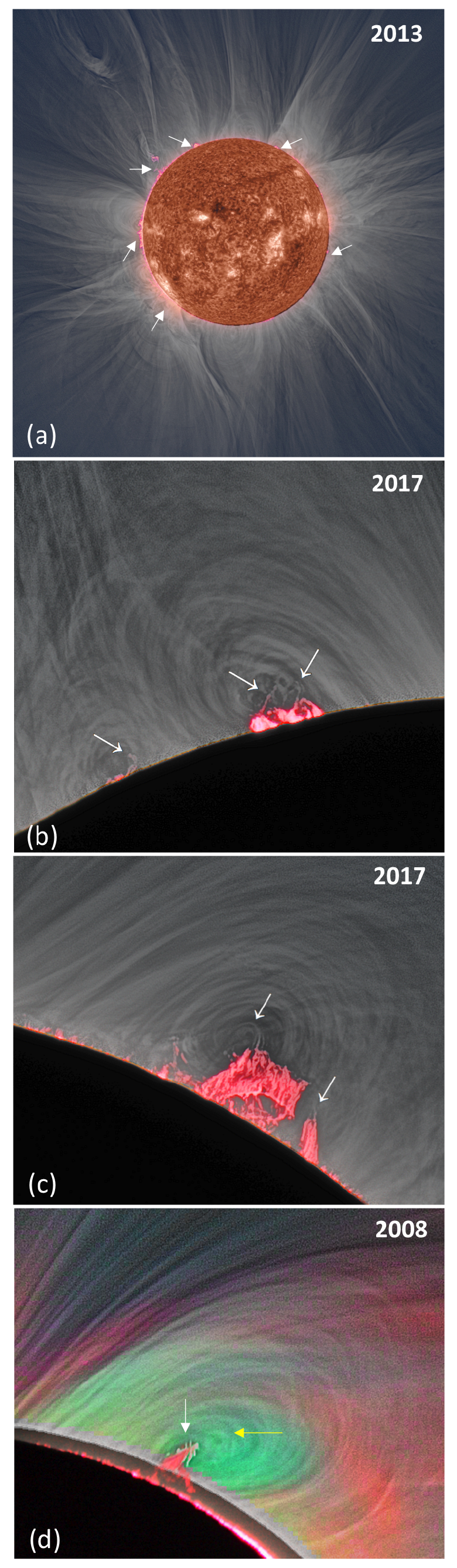}
}
\caption{ Connectivity between prominence structures (pinkish H$\alpha$ emission) and large scale coronal structures. (a) The 42 s 2013 eclipse was such that the Moon's angular extent was almost identical to the Sun's, thus enabling a perfect match between the SDO/AIA He II 304 chromospheric emission showing how all the prominences at the base of the corona around the solar disk, are connected to the fine filamentary structures in the overlying corona.  (b) and (c) show two sections of the 2017  eclipse white light image. The arrows point to the links between filamentary structures in prominences and those in the overlying coronal structures. (d) shows a prominence (white arrow) enshrouded by \ion[Fe xiv] (green) emission (yellow arrow), taken during the 2008 eclipse. Red is \ion[Fe xi] emission. }
\label{prom-link1}
\end{figure}

It is clear from Fig. \ref{allWL} that prominences are intricately connected to dynamic events, as they expand within the core of CMEs, while remaining tethered to the Sun. The examples in Fig. \ref{prom-link1} show that the link between prominences and the surrounding corona is far more ubiquitous. The 2013 example, shown in the top panel, is particularly informative.
This eclipse was special because of its 40 s duration. As such, the angular extent of the Moon almost perfectly matched that of the solar disk, thus enabling coronal structures to be traced all the way down to the solar surface. This further enabled a perfect match with SDO/AIA \ion[He ii] 30.4 nm disk observations. 
The ubiquitous presence of prominences at the base of all streamers (indicated by the white arrows) is especially pronounced at solar maximum when streamers are almost evenly distributed around the solar limb.

Another fine example of the connectivity between prominences and surrounding coronal arches is shown in the details of the 2017 white light image in the middle panels, where the spatially resolved structures within two prominences were captured in two sections of the corona. The arrows point to the unmistakable connectivity between the filamentary structures belonging to a prominence, and those associated with the overlying coronal arches, mentioned earlier. The details in the lower panel, taken from the 2008 eclipse, show how this connectivity reflects a link between the cool prominence material, typical of chromospheric emission, protruding into the corona, and the hot \ion[Fe xiv] emission in the bulge of streamers, referred to as `hot prominence shrouds' by Habbal et al. (2010c, 2014).  This connectivity accounts for the close association of \ion[Fe xiv] emission with CMEs, waves and turbulence in the corona, as noted earlier. 

\subsection{Prominences and the Solar Cycle}

\begin{figure}[ht]
\centering
{\includegraphics[width=0.8\textwidth]{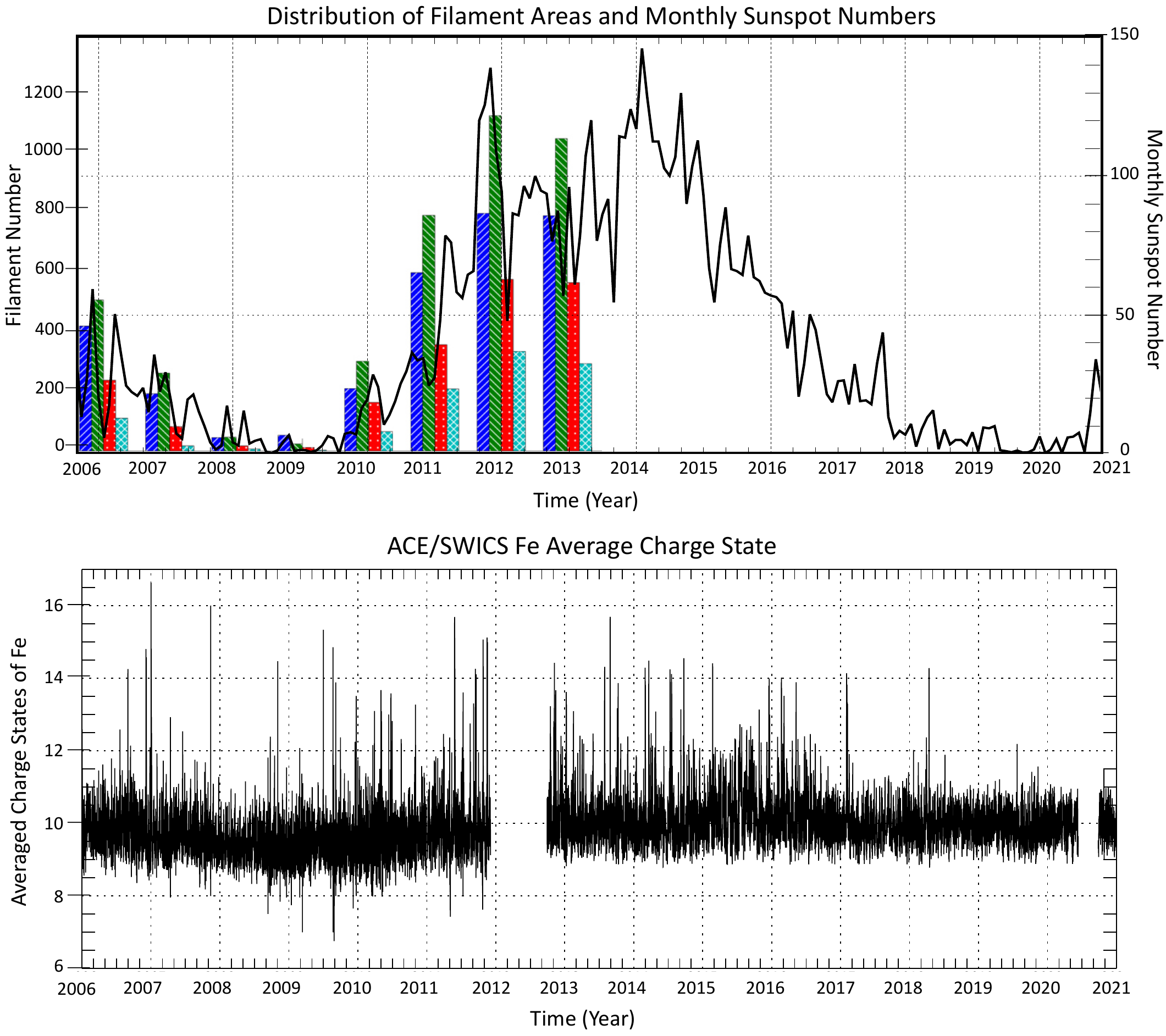}
}
\caption{Top: Overlay of monthly sunspot number and yearly histograms of the areal distribution of prominences on the solar surface up to 2013 (from Hao et al. 2015). The sunspot number extends to 2021 to cover the time span of the eclipse observations. In the histograms, dark blue is for areas $< 2.5 \times 10^8 ~ km^2$, green for the range $2.5 \times 10^8  \--\ 5 \times 10^8 ~ km^2$,  red for the range $5 \times 10^8  \--\ 1.0 \times 10^9 ~ km^2$, and light blue for values $ >  1.0 \times 10^9 ~ km^2$. Bottom: ACE/SWICS 12 minute averaged Fe charge states from 2006 to 2021, plotted every 2 hours.    }
\label{hao-ace}
\end{figure}

The ubiquitous connectivity between prominences and streamers, and the changes of the spatial distribution of the coronal temperature associated with changes in the spatial distribution of streamers as a function of solar cycle, strongly suggest that prominences play an essential role in driving the latter changes. Here we refer to the recent study by
Hao et al. (2015) for additional supporting evidence. These authors explored the areal distribution of prominences across the solar surface using H$\alpha$ data from Big Bear Solar Observatory collected over three cycles between 1988 and 2013. They binned their data into four categories depending on the areal coverage of prominences, as shown by the histograms in the top panel of Fig. \ref{hao-ace}, together with a plot of the sunspot number  
 covering the extent of the eclipse observations. (Unfortunately, the last data included in their study was from 2013.)  A clear solar cycle dependence is evident for all areal coverage bins.  
 
The connectivity between prominences, surrounding coronal arches, and streamers, established from the eclipse observations presented earlier, together with the change of the areal distribution of prominences across the solar surface, as shown by Hao et al. (2015), provide further supporting evidence that prominences play a fundamental underlying role in triggering changes in the topology of the coronal magnetic field, and thus changes in the spatial distribution of temperature in the corona, with solar activity.
  
\pagebreak

\section{Connecting the Corona to the Solar Wind}
 \label{fe_ace}

Shown  in the lower panel of Fig. \ref{hao-ace}, are the ACE Fe charge state measurements, described earlier in section 2.2, where they are referred to as Fe average charge state, since they are the 12-minute averaged measurements plotted every 2 hours.  They encompass the ions responsible for the Fe coronal emission lines, in particular $\rm Fe^{10+}$ for \ion[Fe xi]  and $\rm Fe^{13+}$ for \ion[Fe xiv].  There are two noteworthy trends in this plot: 
(1) The Fe charge states are mostly clustered in a band between $\rm Fe^{9+}$ and $\rm Fe^{11+}$, centered on  $\rm Fe^{10+}$, regardless of time within the solar cycle,
with a slight decrease in 2009, which is statistically insignificant.  
(2) The  peaks in Fe charge states $>$ $\rm Fe^{11+}$ appear sporadically throughout the whole time period.

\begin{figure}
\centering{
\includegraphics[width=0.8\textwidth]{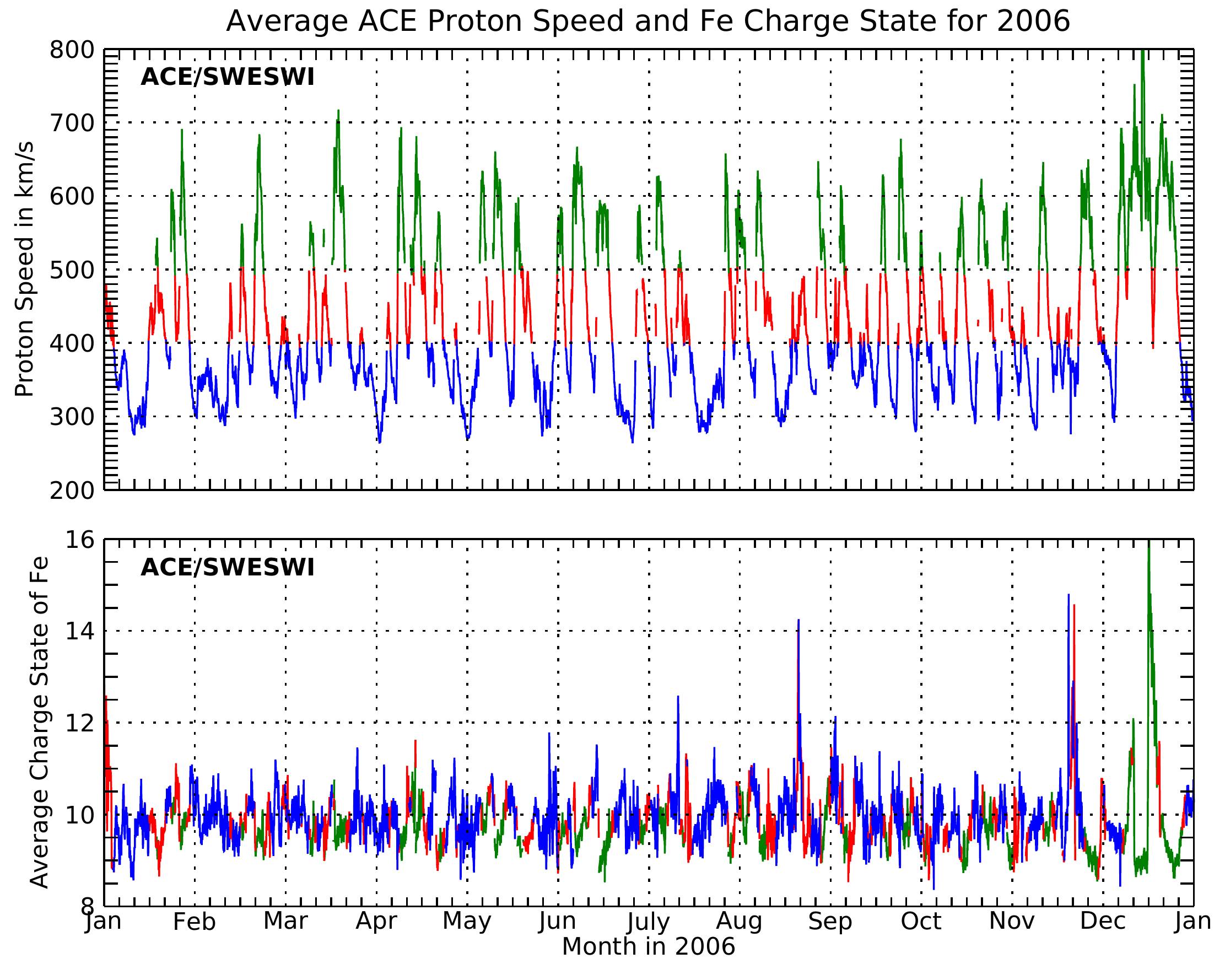}
}
\centering{
\includegraphics[width=0.8\textwidth]{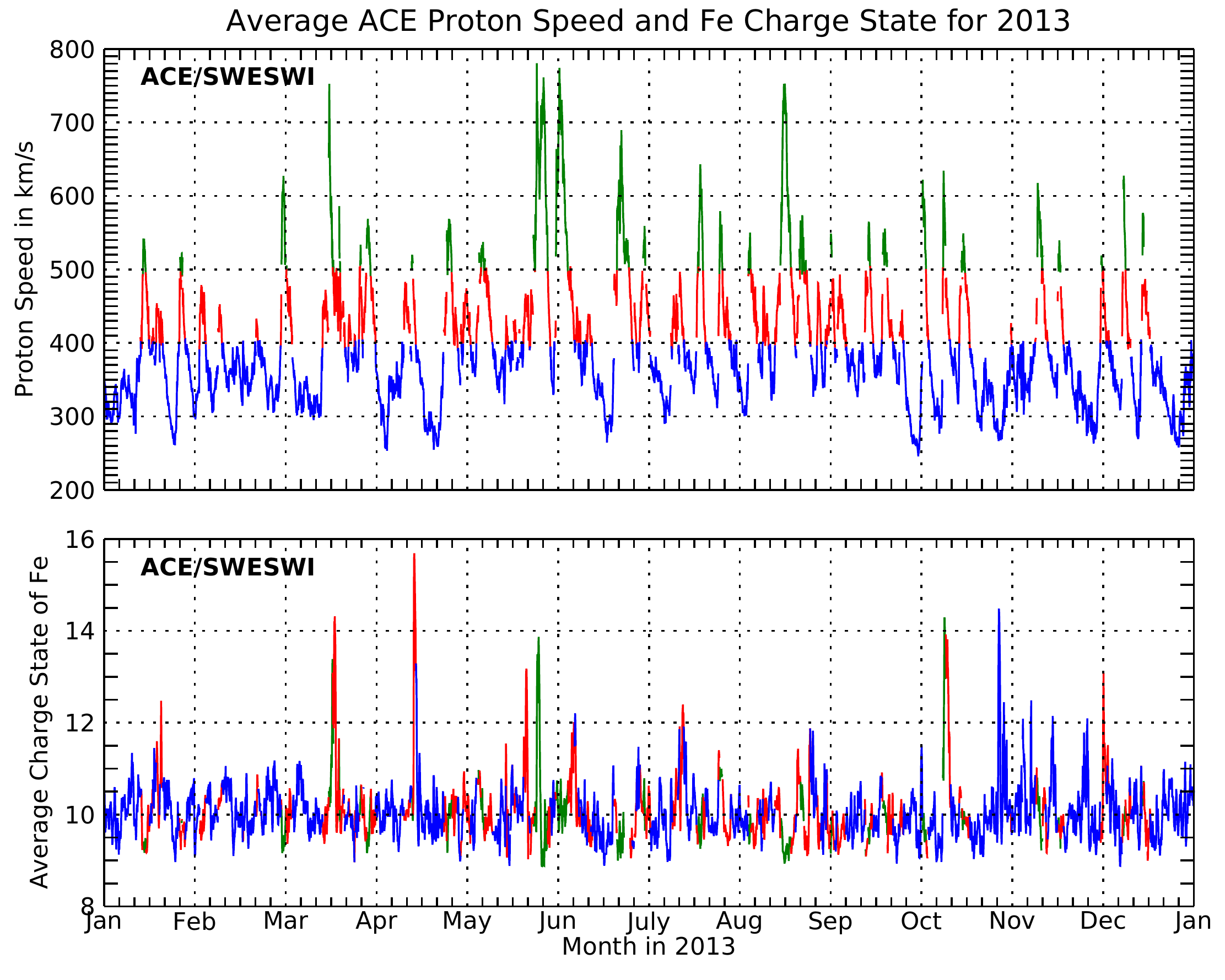}
}
\caption{Two-hour time averaged Fe charge state and corresponding solar wind speed taken from the SWEPAM-SWICS data on ACE, for 12 months in 2006 during the declining phase of SC 23 (top panels), and in 2013 during the peak of SC24 (lower panels). Blue represents data for speeds below 400 $\rm km s^{-1}$,  red for speeds in the range of 400 \--\ 500 $\rm km s^{-1}$, and green for speeds larger than 500 $\rm km s^{-1}$.}
\label{acenew}
\end{figure}

We consider next the solar wind speed data associated with the Fe average charge states. Given the large data volume, we choose two representative time periods: 2006 during the descending phase of SC23,  and 2013 at solar maximum of SC24, as shown in Fig. \ref{acenew}. The data have been color-coded for three speed ranges: blue for speeds $<$ 400 $\rm  km ~ s^{-1}$, red for 400 to 500 $\rm  km ~ s^{-1}$, and green for speeds $> \rm 500 ~ km ~ s^{-1}$. The corresponding charge states are color-coded accordingly.
It is clear from these data that solar wind streams vary continually in speeds between 300 and 700 $\rm km ~ s^{-1}$, throughout the different phases of any given SC, despite the persistent prevalence of $\rm Fe^{10+}$. These streams are thus referred to as the `continual' solar wind.

The freeze-in distance for Fe ions in the corona, inferred empirically by Boe et al. (2018), and from model calculations (e.g. Landi et al. 2012, 2014), shows that ionization equilibrium breaks down and ions become frozen-in within the field of view of the eclipse observations. This implies that the \insitu\ solar wind ion population, dominated by $\rm Fe^{10+}$ ion charge state \insitu, is already determined below a heliocentric distance range which is contained within observations of \ion[Fe xi] emission in the corona. Furthermore,  Figs. \ref{ion}c, \ref{allfe1} and \ref{allfe2} consistently show the prevalence of \ion[Fe xi] ($\rm Fe^{10+}$) emission throughout the corona at any phase of the solar cycle, associated with plasmas at \Txi.  $\rm Fe^{10+}$ thus provides a fiducial link between the continual solar wind streams spanning speeds from 300 to 700 $\rm km ~ s^{-1}$ and their \Txi\  sources in the expanding corona, independently of solar cycle activity.

The sporadic and occasional peaks in Fe charge states, exceeding $\rm Fe^{11+}$, are most likely associated with Interplanetary CMEs (or ICMEs) and CMEs in the corona.  Inspection of the ACE ICME catalog (see Richardson \& Cane 2010, and http://www.srl.caltech.edu/ACE/ASC/DATA/level3/icmetable2.htm) and the LASCO/C2 CME catalogs (https://cdaw.gsfc.nasa.gov/CME$\_$list/) shows that in 2006, an interplanetary disturbance was identified at 14:14 UT on 12/14, and at 17:55 UT on 12/16 (see top panel in Fig. \ref{acenew}). Their most likely sources are a halo CME reported in LASCO C2 at 02:54 UT on 12/13 with a linear speed of 360 $\rm km s^{-1}$ followed by a halo CME at 22:30 UT with a linear speed of 1042 $\rm km s^{-1}$ on 12/14.
In 2013, an interplanetary disturbance with a speed of approximately  $\rm 500 ~ km ~ s^{-1}$ identified in ACE at 22:54 UT on 04/13 (see bottom panel in Fig. \ref{acenew}), originated from a halo CME in the corona at 07:24 UT with a linear speed of  $\rm 861 km ~ s^{-1}$  on 04/11. While the association of CMEs with higher Fe charge states has been known for some time (e.g. Lepri et al. 2001), their drivers at the Sun had not been established. 

Finally, we note that the other peaks in Fe charge states, with relatively low speeds (colored red and blue), in Fig. \ref{acenew}, have no corresponding references to CMEs in the aforementioned ACE and LASCO C2 catalogs. Those are most likely the result of dynamic events releasing hot material from streamers, since they are the only structures identified with the higher temperature \ion[Fe xiv] ($\rm Fe^{13+}$) emission in the expanding corona. Given that prominences are the only dynamic entities identified in the eclipse images directly connected to streamers, they most likely enable the release of  hot streamer material into the solar wind through sporadic magnetic reconnection. 

\section{Discussion} 
\label{discuss}

\subsection{The Constancy of the Electron Temperature at the Source of the Continual Solar Wind}

The composite eclipse images shown in Figs. \ref{ion}c, \ref{allfe1} and \ref{allfe2} provide ample evidence of the manifestation of the different phases of a solar cycle in the distribution of, and changes in streamers. While evidence of the solar cycle in the changing `shape' of the corona has been known from white light eclipse observations dating back to over a century (see, e.g. Hansky et al. 1901, Mitchell 1932), the preponderance of topological open magnetic field structures, originating at the solar surface and expanding out to several solar radii, is far more pronounced in these high resolution images presented here. The ubiquity of topologically open structures, also confirmed by the recent work of Boe et al. (2020b) who applied the rolling Hough transform (RHT) to almost two solar cycles of high spatial resolution white light eclipse images acquired between 2001 and 2019, underscores the view that sources of the expanding corona and the solar wind are not  limited to polar coronal holes, or the boundaries of streamers, as first pointed out by Habbal \& Woo (2001) and Woo \& Habbal (1997).

The direct association between \ion[Fe xi] emission and open magnetic structures further underscores the constancy of the electron temperature at the source of continual solar wind.
The empirical inference of the electron temperature at the sources of the solar wind has been a topic of continued investigation, given its role as an empirical constraint for  models of solar wind acceleration. Earlier inferences from different instruments yielded a range of values (see e.g. Habbal et al. 1993; David et al. 1998). One of the inherent limitations in inferences from remote sensing observations is contamination from different temperature plasmas along the line of sight. Values quoted for coronal holes ranged from 0.7 to 1.6 MK, with a more likely average value below 1.3 MK, and 0.8 - 0.9 MK at their base (see Habbal et al. 1993).  Recently, Morgan \& Taroyan (2017) found that the mean coronal temperature from solar disk observations in the multi-filter bandpasses of the SDO/AIA instrument, spanning 2010 to 2017, which included the peak in Solar Cycle 24, varied from 1.4 to 1.8 MK. However, the AIA/SDO observations have concerning limitations since none of the designed spectral channels at 9.4, 13.1, 17.1, 19.3, 19.5, 30.4 and 35.5 nm, contains emission from a single ion, thus impacting  the reliability of the interpretation of the thermal characteristics of the coronal plasma, as recently noted by Del Zanna (2019). 

However,  inferences of average temperatures do not enable the distinction between the freely streaming coronal plasma, which we refer to as the `expanding' corona, and the coronal plasma bound to the Sun.  By considering total solar eclipse images from coronal emission lines from distinct electron temperature plasmas, straddling two solar cycles, an unequivocal identification of the temperatures associated with the two topologically different magnetic field structures  dominating the corona was achieved.

There is no doubt that the solar wind evolves from its sources at the Sun into interplanetary space, as first comprehensively documented by the Helios measurements (Schwenn 1990). The evolution of the solar wind characteristics with its expansion will continue to be subject of exploration with NASA's Parker Solar Probe and ESA's Solar Orbiter. However, the unique set of total solar eclipse observations straddling more than a full solar cycle, presented here, shows that the dominant presence of  $\rm Fe^{10+}$ in topologically open coronal structures, survives as the dominant ion, \insitu, regardless of the corresponding solar wind speed. This `continuous' link between the corona and the solar wind implies that the `continual' wind expanding from the corona, reaches a range of speeds in interplanetary space, depending on the physical parameters at their origin. Given that the dominance of \ion[Fe xi] emission in the expanding corona implies an approximately constant  \Txi\ there, leaves the density at the coronal base of solar wind streams as the  main parameter that determines the asymptotic solar wind speed. This is indeed confirmed by earlier solar wind models by Leer \& Holzer (1980) and Habbal et al. (1995). Simply stated, the energy input to higher density wind sources at the Sun ends up being  distributed among a larger number of particles, more so than for lower density regions, thus leading to slower streams in the former, and faster in the latter. Furthermore, the coronal conditions determining this continual wind imply that the same physical process in the corona drives the different speed streams. This yet-to-be-identified physical mechanism caps the electron temperature in the expanding corona at an almost constant value, \Txi, namely the peak ionization temperature of $\rm Fe^{10+}$.

We caution that  the  \Txi\ associated with \ion[Fe xi] emission at the sources of the solar wind outflow, whether fast or slow, is probably uncertain to within $\approx$ 10 \--\ 20 \%, based on different published works (e.g. Del Zanna \& De Luca 2018, Landi, private communication).  Indeed, the peak ionization temperatures presented in Fig. \ref{ion} are subject to the assumption of ionization equilibrium, which is not necessarily valid in the corona.  Regardless, the observations of \ion[Fe xi] and \ion[Fe xiv] emission in the corona indicate that the relative spatial distribution of the two charge states is the same observable seen \insitu. Thus, even if the precise temperature values are not correct, the \ion[Fe xi] and \ion[Fe xiv]  frozen-in charge states quantitatively link solar wind types to their sources in the corona, as demonstrated here. In a sense, the temperature values are just a convenient number for understanding the observed charge ratio. The main result is that the constancy of \Txi\ in the expanding corona, rather than its exact value, leads to the continual solar wind with a range of speeds.

\subsection{Prominences, CMEs and the Variable Solar Wind}

The ubiquitous presence of prominences at the base of streamers, their unmistakable connectivity to the overlying `closed' coronal structures (see also  (Habbal et al. 2010c, 2014; Druckm\"uller et al. 2017), and the finding that streamers are the hotter structures in the expanding corona, also emerged from these unique eclipse observations.  In fact, the most comprehensive EUV space-based data from AIA/SDO lack the spatial extent and adequate temperature diagnostic to reach such a conclusion. This limitation is particularly critical for understanding the thermal properties of CMEs, which originate within the bulges of streamers. 
It is through this link that the coolest material protruding into the corona, typical of chromospheric emission from neutral or low ionized atoms,  finds itself enshrouded by the hotter coronal plasma (see Habbal et al. 2010c, Habbal et al. 2014).  

Prominences are known to be the most dynamic structures in the corona, with their frequent eruptions being directly linked to  CMEs. As shown in Figs. \ref{allWL} and \ref{prom-link1}, the close link between the ejected prominence and coronal material forming a CME front is a direct consequence of the link between the two. The invariable association of ICMEs with high charge states \insitu\ is further confirmation of this close link. Another signature of the consequence of prominence dynamics in \insitu\ measurements  is the detection of neutrals and low ionized atoms, characteristic of prominence plasmas (Gloeckler et al. 1998, Lepri et al. 2013). Indeed, 
the fate of eruptive prominence material in conjunction with a CME was recently discovered in spectroscopic observations by Ding \& Habbal (2017) during the 2015 total solar eclipse. These authors captured Doppler red-shifted emission in \ion[Fe xiv], with speeds ranging from 100 to 1500 km $\rm s^{-1}$, associated in about 10\% of the time with emission from neutrals and low ionized atoms, characteristic of prominence material. These observations were the first to yield direct evidence for the escape of erupting prominence material, unscathed into interplanetary space. Magnetic reconnection events within the prominence/corona interface remains an essential process for the release of the two extreme components, namely, neutrals and low ionized atomic states, and the highest charge states from bound structures in the corona. The variable solar wind, likely to also be turbulent, as visually present in the eclipse images,  is then distinctly different from the continual solar wind.  It is likely to be associated with the sporadic presence of two extreme charge states: neutrals and/or low ionized charge states originating from ejected prominence material, and charge states exceeding ${\rm Fe}^{13+}$, but not necessarily producing CMEs. 

\section{Conclusions}
 \label{concl}
Images of the  corona acquired simultaneously in white light, \ion[Fe xi] and \ion[Fe xiv], during total solar eclipses between 2006 and 2020, thus spanning more than an entire solar cycle, are presented here for the first time.  Complemented by ACE \insitu\ measurements of Fe charge states and solar proton speeds spanning the same time period, they yield novel insights into the coronal sources of the solar wind.  Given that the $\rm Fe^{10+}$ and $\rm Fe^{13+}$ charge states, corresponding respectively to \ion[Fe xi] and \ion[Fe xiv] emission in the corona, are frozen-in within the field of view of the eclipse images, their spatial distribution in the corona is then directly correlated with their corresponding charge state distribution \insitu.
In particular, this data complement points to the presence of  continual solar wind  streams,  dominated by  \ion[Fe xi] emission with a temperature \Txi\ = 1.2 $\pm 0.1$ MK in the expanding corona,  with  \insitu\ Fe charge states clustering around $\rm Fe^{10+}$, and speeds ranging from $\approx$ 300 to 700 $\rm km ~s^{-1}$, throughout the different phases of a solar cycle.   The spatial distribution of \ion[Fe xiv] emission and  $\rm Fe^{13+}$ charge states exhibit large variances with solar cycle, indicating that the hotter corona and corresponding wind depend on large-scale coronal structures, notably streamers, changing with the solar cycle. 

This complement of coronal and \insitu\ measurements also connects  dynamic events in interplanetary space to the intrinsic dynamics of prominences,  which are invariably linked to coronal arches forming the base of streamers, as well as active regions. Consequently, inevitable magnetic reconnection events within the complex structures of prominences and their surrounding regions, lead to the expulsion of two extreme populations of plasma composition: neutrals and/or low ionized ions, and Fe charge states exceeding $\rm Fe^{11+}$,  detected \insitu. The high ionized charge states are linked to the confined high temperature plasmas released from the bulges of streamers, as a direct consequence of prominence dynamics, whether through the formation of CMEs, or the release of a turbulent solar wind. The low charge states are remnants of expelled prominence material with typical chromospheric composition.

While the impact of the different phases of a solar cycle is evident in the changing topology of streamers, it does not seem to have any effect on the unidentified physical processes which maintain the electron temperature at the sources of the continual outflow from the Sun at an almost constant value.  These findings thus yield new constraints on models of the solar wind.

\acknowledgments

S. R. Habbal and the Solar Wind Sherpas acknowledge support from NASA and NSF  for the eclipse observations spanning more than a solar cycle; the most recents grants being NASA Grant NNX08AQ29G and NSF grant ATM 08-02520 to the University of Hawaii.  N. Alzate acknowledges support from NASA through an appointment to the NASA Postdoctoral Program at the Goddard Space Flight Center, administered by Universities Space Research Association under contract with NASA, and support under HGI Grant No. 80NSSC20K1070.  B. Boe was supported by the National Science Foundation under Award No. 2028173. M. Druckm\"uller was supported by Grant Agency of Brno University of Technology, grant FSI-S14-2290. M. Arndt's contribution to the eclipse observations was supported by Bridgewater State University CARS grants. We would also like to acknowledge valuable discussions with C. Haggerty, E. Landi, J. Scudder and M. Velli, as well as the insightful input from the anonymous reviewer. 


\end{document}